\documentstyle[preprint,floats,aps,epsf,epsfig,prb]{revtex}
\tightenlines

\newcommand \ltdash{\raise-1.8pt\hbox{$\scriptscriptstyle |$}}

\newcommand \bea {\begin{eqnarray} }
\newcommand \eea {\end{eqnarray}}

\newcommand \rarrow{\rightarrow}
\newlength{\bxwidth}\bxwidth=0.8\textwidth

\newcommand\prm[2]{$ $\vskip 2.4 truein
\centerline{\epsfig{file=#1,width=\bxwidth} }\vskip 0.5truein
\centerline{{\bf Fig.} #2}}
\newcommand\prk[2]{$ $\vskip 0.5 truein
\hskip 1.  truein
\epsfig{file=#1,width=\bxwidth} \vskip 0.3 truein
\centerline{{\bf Fig.} #2}
}
\begin{document}
\draft
\title{Onset of Antiferromagnetism in heavy fermion metals.}
\author{A. Schr{\"o}der$^1$, G. Aeppli$^2$,
R. Coldea$^{3,4}$, M. Adams$^4$, O. Stockert$^{1,5}$, H.v.
L{\"o}hneysen$^1$,
E. Bucher$^{6,7}$, R. Ramazashvili$^8$ and P. Coleman$^9$
}
\address{$^{1}$ Physikalisches Institut, Universit{\"a}t
Karlsruhe, D-76128 Karlsruhe, Germany}
\address{$^{2}$ NEC, 4 Independence Way, Princeton, NJ 08540, U.S.A.}
\address{$^{3}$ Oak Ridge National Laboratory, Oak Ridge, TN 37831,
U.S.A.}
\address{$^{4}$ ISIS Facility, CCLRC, Rutherford- Appleton Laboratory,
Didcot OX11 0QX, UK}
\address{$^{5}$ present address: University of Bristol, Bristol BS8 1TL,
UK}
\address{$^{6}$ Universit{\"a}t Konstanz, D-78457 Konstanz, Germany}
\address{$^{7}$ Bell Laboratories, Lucent Technologies, Murray Hill, NJ
07974, U.S.A.}
\address{$^{8}$Department of Physics, University of Illinois, Urbana,
IL
61801, U.S.A.}
\address{$^{9}$ Materials Theory Group,
Rutgers University, Piscataway, NJ 08855, U.S.A. }
\maketitle
\vskip 1truein

{\bf 
There are two views of antiferromagnets. The first
proceeds from atomic physics, which predicts that atoms with unpaired
electrons develop magnetic moments. In a solid, the coupling between moments on nearby ions then yields antiferromagnetic order at low
temperatures \cite{anderson1}. The second, based on the physics of
electron fluids or 'Fermi liquids', states that Coulomb interactions can drive  the
fluid to adopt a more stable configuration by developing a
spin density wave.  \cite{stoner,overhauser} It is presently unknown
which view is appropriate at a `quantum critical point', where the
antiferromagnetic transition temperature vanishes
\cite{hertz,millis,chakravarty,sachdevscal}. Here we describe 
 an atomically local contribution to the magnetic
correlations which develops in the 
metal CeCu$_{6-x}$Au$_{x}$ at the critical gold concentration
($x_c=0.1$) where the magnetic ordering temperature is tuned to zero.
This contribution implies that a Fermi-liquid destroying spin-localizing transition, unanticipated
for the spin density wave description, 
coincides with the antiferromagnetic quantum critical point

}

CeCu$_6$ \cite{stewart,onuki} is a ``heavy fermion'' compound, a class
of metal formed between actinide or rare earth elements and transition
or noble metals. 'Heavy' refers to the extremely large effective
masses of the charge carriers at low temperatures, often hundreds or
even thousands of times greater than the electron mass.  
Because of their large effective electron masses and proximity to
antiferromagnetism\cite{doniach,broholm}, heavy fermion materials are ideal venues for
observing the competition between metallic electron band formation and
magnetism. The large masses imply small bandwidths, and the magnetism is
 easily tuned by external pressure \cite{mathur}, magnetic field and 
chemical
composition. Thus, all of the phenomena associated with the competition
can be readily accessed at low temperatures where it is much easier to
collect clean data than for other materials, such as the transition
metals or their oxides \cite{varma,sachdevscal}, 
which also display competition between band
formation and magnetism but whose characteristic temperatures are much
higher. Because of a substantial data base \cite{cca,osto,rho,early,ccath} and the
ability to grow large and highly homogeneous single crystals,
we have chosen CeCu$_{6-x}$Au$_{x}$ as a test material. The quantum critical
point (QCP) occurs as gold is substituted for the Cu atoms; when more
than 0.1 Cu sites per cerium are replaced\cite{ruck}, the heavy
fermion paramagnet gives way to an ordered antiferromagnet
\cite{schroder,stockert} (Fig.\ref{1}a).

Fig.\ref{1} illustrates two competing models of magnetic order in
rare earth metals that have been widely discussed in the context of
heavy fermion systems\cite{si,rosch,doniach88,japanese}.  The
fundamental parameter is the strength of the hybridization $W$ between
the localized f orbitals and the more extended s, p and d
orbitals. When $W$ is small, the extended orbitals form a metal which
is weakly coupled to the unfilled, and hence magnetic, f-orbitals. Thus, as $T$ is reduced,  the two subsystems
undergo largely independent evolutions 
towards Fermi liquid and
magnetically ordered ground states, a situation found in many
elemental rare earths(e.g. gadolinium), and their
compounds.  The volume bounded by the Fermi surface is ``small'',
containing only electrons
from the extended orbitals.   As we begin to increase $W$, the magnetic
coupling between the rare earth ions via the conduction electrons and
the magnetic ordering (or ``N{\'e}el'') temperature, $T_N$, both
increase.
On the other hand, if $W$ is large, the material develops a Fermi
liquid of strongly hybridized carriers with a ``large'' Fermi volume 
containing electrons from
both extended and f orbitals.  The kinetic energy of
band electrons then overwhelms interaction effects, with the result
that $T_N$ is suppressed to zero for $W$ beyond a critical value
$W_{c}$ \cite{doniach}.

A key scale  for heavy fermion materials is  the ``Kondo temperature''
$T_K$:  the  temperature   below  which  the  magnetic  susceptibility
saturates  and the  hybridization  of the  f-orbitals  with the  other
orbitals becomes apparent. In contrast to $T_N$, the Kondo temperature
$T_K$, increases  monotonically with  $W$.  The central  open question
concerns how  this scale behaves  near the quantum critical  point: in
particular, whether  it remains finite,  (Fig. 1b) or  whether $T_{K}$
and $T_{N}$ vanish  at the same point(Fig. 1c).  The first scenario is
required if  the local  moments are quenched  at a  finite temperature
above the  quantum critical point: here,  local moments do  not play a
role in the  physics of the quantum critical  point and a spin-density
wave  model must be  used.\cite{millis,rosch,doniach88} In  the second
case, local moments  are present at all temperatures  down to $T=0$ at
the QCP,  so a  heavy electron  Fermi surface ceases  to exist  at the
quantum   critical  point.    The  present   paper  gives   
comprehensive   evidence   in    favor   of   the   second   scenario,
fig.\ref{1}(c).

To discriminate between the local-moment and spin-density-wave visions,
we have measured the magnetic response function
$\chi({\bf
q},E)$, as a function of temperature (T) and external magnetic fields
(H), using neutron scattering and bulk magnetometry. The function
$\chi({\bf q},E)$ describes the Fourier components at frequencies
$E/\hbar $ of the magnetization $\delta m({\bf q},t)= \chi({\bf q},t)
\delta h_o$
induced by a unit external field pulse  $\delta h({\bf r},t)=\delta h_0
e^{i{\bf q}\cdot {\bf r}}\delta(t)$
with spatial modulation described by the wavevector $\bf q$.
A particularly simple impulse
response is an exponential decay,
$\chi({\bf q},t)\propto exp[\frac{-\Gamma_{\bf q}}{\hbar}t]$, where
$\Gamma_{\bf q}/\hbar$ is the magnetic decay rate,
for which
\bea
\label{eq1}
\chi({\bf q},E)=A/(\Gamma_{\bf q}-iE),
\eea
where $A$ is a constant.
The exponential form is typical
for paramagnetic rare-earth insulators
where the  magnetization is not conserved.
To understand how $\Gamma_{\bf q} $ can depend on temperature,
recall that
the zero-energy response is the static magnetic susceptibility,
which
in paramagnetic insulators
follows a Curie-Weiss law $ \chi(
{\bf q},E)|_{E=0} =C/(T+\theta ({\bf q}) )$ with Curie constant $C$
and {\bf q}-dependent Weiss temperature $\theta({\bf q})$. At ${\bf q}=0$
this is just the uniform susceptibility, which can be
directly measured using a magnetometer.
Comparison
with eq. (\ref{eq1}) leads to the conclusion that
$\Gamma_{\bf q}=a(T+\theta (\bf q))$ where
$a=A/C $. Hence,
in paramagnetic insulators, the inverse magnetic response function
\bea
\label{loc}
\chi^{-1}({\bf q},E,T) = \chi _{0}^{-1}(E,T)+\theta^{\alpha}({\bf  q})/C
\eea
is a sum of a non-local, purely  $\bf  q$-dependent piece
$\theta^{\alpha} ({\bf q})/C$  (here with $\alpha=1$) and a local 
($\bf  q$-independent) driving
term $\chi _{0}^{-1}= ( T-iE/a)/C$. Thus $T$ and $iE/a$ are
interchangeable in
expressions for $\chi ({\bf  q},E)$, and at critical wavevectors, i.e.
the points in the Brillouin
zone where $\theta ( {\bf  q})=0$, $\chi=\chi_0$
exhibits $E/T$ scaling, meaning that
\bea
\label{et}
\chi_0  =T^{-\alpha} g(E/T)=E^{-\alpha} j(T/E)
\eea
where $j(1/y)=y^{\alpha} g(y)$.   In the simple Curie-Weiss example,
$\alpha=1$ and $g(y)=g_{0}(y)=C/(1-iy)$ with $y=E/aT$.

While
$E/T$-scaling at critical wavevectors is well-known and
established\cite{sachdevscal,varma,aronson,early}, 
it does not capture the general dependence of $\chi$ on
wavevector, magnetic field, and zero-temperature tuning parameter (e.g.
pressure or composition) for the quantum phase transition. However,
inspection of the Curie-Weiss example suggests a set of hitherto
untested scaling laws and data replotting procedures. The first 
follows from insertion of the $T\rightarrow 0$ limit of eq.(\ref{et}) 
into eq. (\ref{loc})
. The outcome is that $\chi^{-1}({\bf q},E,T=0)$
displays $E/\theta({\bf q})$ scaling, i.e.
\begin{equation}
\label{eth}
\chi(T\rightarrow 0)=\theta({\bf q})^{-\alpha} G(E/\theta({\bf q}))
\end{equation}
In the simple Curie-Weiss case 
$G(\delta)=g(\delta)$ (with $\delta=E/a\theta$),
which is easily understood if
we regard excursions in wavenumber or quantum control parameter as
excursions in an effective Curie temperature. Letting $E\rightarrow 0$ instead
of $T\rightarrow 0$ 
in eq.(\ref{et}) and again inserting the result into eq.(\ref{loc})
gives a second verifiable relation,
\begin{equation}\label{e0}
\chi^{-1} ({\bf q},T)= (T^{\alpha}+\theta ({\bf  q})^{\alpha})/C,
\end{equation}
Eq.(\ref{e0}) implies that the traditional Curie-Weiss plot, where
$1/\chi$ is a straight line as a function of $T$, can be replaced across
a series of experiments where the Weiss temperature is varied (e.g. by
varying ${\bf q}$ or composition $x$) by a plot where $1/\chi$ as a 
function of
$T^\alpha$ is a straight line with intercept $\theta ({\bf q})^{\alpha}$.

Finally, combining the knowledge that for local moment systems, the
magnetic 
response is
governed by the ratio of the Zeeman energy $g\mu_BH$ to the thermal
energy
$k_BT$, with eq.(\ref{loc}) yields another scaling law, namely
\begin{equation}\label{ht}
(\chi(H,T)^{-1}-\theta(0)^{\alpha}/C)^{-1}T^{\alpha}=f(H/T)
\end{equation}
The law eq.(\ref{ht}) can be tested with extraordinary sensitivity in the
long wavelength $(q=0)$  limit using bulk magnetometry.  

For systems in which the critical magnetic degrees of freedom derive
from spin-density fluctuations of a Fermi sea, matters are much more
complicated and do not satisfy eqs(\ref{loc})-(\ref{ht}).  Here we
expect that non-local terms, derived from the band-structure, should
have both $E-$ and $T-$ dependence. At low $T$, the first
correction to the $T=0$ Pauli susceptibility is of order $T^2$ rather
than $T^{\alpha}$, as suggested by the generalized Curie-Weiss
Ansatz. Also, $T$ and $iE$ are not interchangeable in the
temperature-dependent susceptibility. In particular, 
above two dimensions,\cite{hertz} the quantum critical fluctuations
of a spin density wave are non-interacting at low energies and long
wavelengths, in which case $E/T$ scaling\cite{sachdevscal}  {\sl
never} occurs.  Likewise, none of the less familiar
forms(\ref{eth})-(\ref{ht}) accounting for the ${\bf q}$-, $T$-, $x$-,
and $H$-dependence should obtain.  Of special significance is that an
external magnetic field modifies the band structure, implying an
influence on the non-local terms and an inability to scale the
field-dependent effects simply in terms of the ratio
$g\mu_{B}H/k_{B}T$.\cite{ht}

We have thus reduced the problem of discriminating between the scenarios
in figs.\ref{1}b) and c) to discovering whether the largely untested laws 
(\ref{eth})-(\ref{ht})
describe the $H$-, $T$-, $E$-, and $x$- dependent magnetic response of 
CeCu$_{6-x}$Au$_x$, and to
verifying that with an order of magnitude improvement in energy range,
$E/T$- scaling obtains at the critical composition and wavevector. To carry
out this ambitious program, we have taken advantage of a low-T dc
magnetometer\cite{hopser} and the 
IRIS spectrometer\cite{carlile}
at the ISIS spallation source. IRIS
represents a substantial improvement over the 
instruments previously used to examine the magnetic fluctuations in
CeCu$_{5.9}$Au$_{0.1}$ in that it (a) surveys 
the magnetic fluctuation spectrum
over a very wide range of wavenumbers, (b) has an energy resolution $9\mu
{\rm eV}\equiv 0.1 K $ almost an order of magnitude better than that of
previous experiments, and (c) yields data in
absolute units established using the simultaneously measured elastic
incoherent scattering from the sample.

As $x$ is varied in $CeCu_{6-x}Au_x$,  Bragg reflections due to
antiferromagnetism appear at specific wavenumbers on a
``butterfly''-shaped critical line \cite{rho},
(black crosses in fig \ref{2}a).
At the quantum critical concentration
$x=x_c$, the Bragg peaks are replaced by magnetic critical scattering -
peaked on the same
``butterfly''\cite{osto}.
Fig. \ref{2}b) and c) show
scans through the butterfly, obtained for the very small
energy
$E=35\mu eV$.
On warming from 50 mK to
$0.4K\approx E/k_B$, the magnetic response at the butterfly is
somewhat reduced, while on going to $1.5K=4E/k_B$, it is greatly
suppressed, showing
that when T crosses $E/k_B$, the magnetic
response begins to change rapidly. Fig. \ref{2}a shows 
energy scans
collected at 50mK at the various (nearly) fixed momentum transfers indicated
in the inset and plotted following the new scaling paradigm eq(\ref{eth}), 
using
the exponent $\alpha \approx 0.75$ obtained from the $E/T$ scaling found in
our previous critical wavenumber $({\bf q}=(1.2,0,0))$ experiment \cite{early} and verified
with much greater confidence below. The data collapse onto a single
curve is clearly consistent with
eq(\ref{eth}), but there is sufficient scatter to mandate checks of the other
untested laws (\ref{e0}) before we can claim to have
verified the local moment scenario giving rise to eq(\ref{loc}). 

Fig.\ref{3} shows the
modified Curie-Weiss plots suggested by eq(\ref{e0}) for 
CeCu$_{6-x}$Au$_{x}$, where the
effective Weiss temperature is varied either by changing $q$ in the
neutron scattering experiment, or by changing $x$ in the bulk (low field)
magnetometry. Near the quantum critical point, the
static susceptibility 
follows the modified Curie-Weiss law
over
the entire two-decade
temperature range of our measurements. As $x$ moves away from $x_c$
in either direction,
$\chi^{-1} (T) $ becomes more curved as $T\rarrow 0$, although there
still is
a quantum critical regime bounded by a cross-over temperature which
increases with distance from $x_c$ and above which the $T^\alpha$ power
law seems to hold (Fig. \ref{3}b), and below which a tendency towards less
singular (heavy) Fermi liquid behavior is seen. Fig.\ref{3}a also 
demonstrates
that at quantum criticality (i.e. $x\approx 0.1$) the same law is 
followed for general $\bf{q}$, where $\chi'({\bf q})$ is obtained via
the Kramers-Kronig relation from the neutron data. Indeed,
the parallelism of the lines in Fig.\ref{3}a shows not
only that a single exponent $\alpha$ is sufficient to account for all
of the magnetometry and neutron data, but also that the Curie constant $C$
is - within the limited statistical accuracy of the latter - the same
for all wavenumbers.  The overlap between the neutron data at
${\bf q}=(1.8,0,0)$ far
from the butterfly but close to (2,0,0), equivalent to (0,0,0),
and the ${\bf q}=0$ bulk results is reassuring and
shows that ${\bf q}=0$ is a
typical non-critical wavenumber, with a spectrum likely to resemble
that measured at ${\bf q}=(1.8,0,0)$. A final result which can be
extracted
from Fig.\ref{3} is another estimate of the wavenumber-dependent Weiss
temperature
$\theta({\bf q})$, which is simply the $x$-axis intercept, taken to the
power $1/\alpha$, of the lines going through the
data for each {\bf q}. The outcome is entirely consistent with that of
the 
$E/\theta({\bf q})$ scaling analysis in fig.\ref{2}a). 

We finally turn to the energy- and T-dependent spectroscopies with the
best signal  to noise ratios,  namely the neutron measurements  of the
scattering  with ${\bf  q}$ on  the butterfly  and the  $H$- dependent
static magnetization.  Fig \ref{4}a  shows the former,  plotted versus
$E/T$,  collected at  a wavenumber  (0.8,0,0).  This  data  set, which
densely covers  four decades  of $E/T$, confirms  that the  scaling is
optimal  (see  left  inset)   for  $\alpha=0.75\pm  0.05$.  
Fig\ref{4}b shows the collapse
of the  bulk magnetization data as  suggested by the  H/T scaling form
(\ref{ht}).  The similarities  in  the scaling  function and  exponent
($\alpha=3/4$), obtained in this  entirely independent assay, to those
obtained  via  neutron  measurements  demonstrates that  the
magnetic response is driven by  a singular local term. Two interesting
details emerge.   First, the effective moment (for  definition see Fig
4)  $\mu=g\mu_{B}=1.5 \mu_B$  is the  same order  as an  atomic moment
($0.6 \mu_B$)estimated  from considering the $q$-space  average of the
neutron data  integrated to $E=1meV$.  This rules  out large, randomly
occurring  ferromagnetic clusters  are a  key feature  of  the quantum
critical  point.\cite{castro}  Second,  the  scaling  function  $f(h)$
(where $h=g\mu_BH/k_BT$) has  the power-law asymptote $h^{-\alpha}$ at
large $h$, which decays much  more slowly than the $exp(-h)$ asymptote
derived  from  the Brillouin  function.  Thus,  the active  ingredient
responsible for the critical  behavior is an atomically local magnetic
moment, with a size comparable with a single spin $S=1/2$ with an
intrinsically critical response to an external field.

Our experiments reveal that the magnetic fluctuations in the
heavy fermion metal CeCu$_{6-x}$Au$_x$ are extraordinarily simple near
the
quantum critical point at $x=0.1$. Not only do our new high-resolution
neutron data satisfy E/T scaling at the critical wavenumbers with much
greater certainty than before, but they together with magnetization data
satisfy two hitherto untested scaling laws and a generalized Curie-Weiss
paradigm, all of which emerge from a local moment scenario for the
quantum critical point.  Detailed quantitative analysis (see figure
captions) reveals that an algebraic form,
\begin{equation}
\chi^{-1}( { \bf q}, E, T) = \biggl[
\biggl (\bigl[T^{2}+({\scriptstyle  \frac{g\mu
_{B}}{k_{B}}})^{2}H^2\bigr ]^{1/2}-iE/a\biggr
)^{\alpha}
+\theta({\bf q})^{\alpha }
\biggr ]
C^{{-1}}=\chi_0^{-1}(E,T)+\theta({\bf q})^{\alpha}/C
\label{lab2}
\end{equation}
accounts for all of our $H$- ${\bf q}$- and $T$- dependent results. 
Eq(\ref{lab2}) is the most straightforward generalization of the Curie-Weiss
description of a critical point and has  only three inputs, an overall
amplitude, an exponent $\alpha$ and the
function $\theta({\bf q})$.

Clearly then, in
answer to the question
posed in the introduction,
whether an itinerant (fig.\ref{1}b) or local (fig.\ref{1}c)
picture of the magnetic quantum critical point applies to
this heavy fermion system, it is
the local scenario (fig.\ref{1}c)  that is appropriate.
Beyond this, the most striking feature of the results is the appearance of a single
unusual exponent $\alpha = 0.75$ in the scaling properties at widely
different wave-vectors.
The
$E/T$ and $H/T$ scaling with the unusual exponent
$\alpha=0.75<1$ means
that the magnetic properties of  CeCu$_{6-x}$Au$_{x}$
are
more severely non-analytic  than those of both Fermi liquids and conventional
insulating magnets, where $\alpha=1$.  

We conclude that a Fermi
liquid, CeCu$_{6}$, with one of the largest known effective carrier
masses is remarkably close to an instability marked by
a new kind of critical behavior where
the moments associated with the tightly bound f
electrons are on the brink of separation
from the conduction electrons- how close is apparent from the approach
of the inverse magnetic susceptibility to the anomalous $T^\alpha$
asymptote as $T$ rises above 2K( fig.3b).
The locality of the quantum fluctuations in the ordered
phase suggests the development of a new kind of localized
excitation, one
derived from the
ambiguity of whether the f-electrons are counted within
the Fermi sea or not.
These  excitations
are neither heavy quasi-particles nor conventional
spin excitations and might best be
regarded as Fermi
surface ``shredders'': new degrees of freedom which are
ultimately responsible for the non-Fermi liquid behavior and the unusual
local scaling behavior at the
quantum critical point.

\vfill \eject

\acknowledgements
A. S., O. S. and H. v. L. acknowledge financial support from the Deutsche
Forschungsgemeinschaft.
R. C. acknowledges
financial support from the Oak Ridge National Laboratory Postdoctoral
Research Associates Program administered jointly by the Oak Ridge
National Laboratory and the Oak Ridge Institute for Science and
Education. P. C. and R. R.  acknowledge the support of the National
Science Foundation.

\newpage

\noindent{\bf Figure Captions}

\renewcommand{\labelenumi}{{\bf Fig.} \theenumi .}
\begin{enumerate}

\item
{\bf  Phase diagram and scenarios for quantum criticality in
$CeCu_{{6-x}}Au_{x}$}
a) Phase diagram for CeCu$_{6-x}$Au$_{x}$, where increasing
Au-concentration $x$ drives the heavy fermion alloy into an
antiferromagnetically ordered state. The  quantum critical
point at $x=0.1$ is the subject of this paper.
b)-c) Two competing models for the magnetic quantum critical point in
heavy fermions. $W$  is
the strength of the hybridization between the localized
f-electrons and the surrounding conduction sea, which causes the
antiferromagnetic order (with transition temperature $T_N$) to
collapse at a critical value $W=W_c$.  (b) The local moments are
magnetically
quenched - absorbed into the metallic bands- at a finite "Kondo"
temperature, $T_K$ and do not
play a role at the quantum critical point. Antiferromagnetism develops
via a ``spin-density-wave instability'' of the underlying Fermi
surface.  
(c) Local moments exist down to an
effective lattice Kondo temperature, $T_K^*$, which vanishes at the
quantum critical point. In this model, local moments become quenched
precisely at the quantum critical point, and play an active role in the
magnetic
critical fluctuations.
\label{1}

\item {\bf  Magnetic neutron scattering data.}
{\bf  (a)} Spectra at wavevectors {\bf q} indicated in inset, as
a function of energy transfer E in units of the {\bf q}-dependent Weiss
temperature  $\theta({\bf q})$. Intensities
are correspondingly normalized to 
$\theta({\bf  q})^{-\alpha}$ (${\alpha}=0.75$), as suggested
by eq.(\ref{eth}).  The line is the 
scaling function $G(\delta)=C/(1+(-i\delta)^{\alpha})$ with $\delta=E/(aT)$.  
The inset is a map of reciprocal space showing the critical line (black lines)
close to the magnetic Bragg peaks in the ordered regime (black crosses)
and the considered wave vector regions (labeled A-H). 
{\bf  b,c,} Wavevector ({\bf q}) dependence of the magnetic fluctuations 
along the middle ({\bf b}) and lower ({\bf c}) trajectory in the inset of
({\bf a}) for fixed
$E=0.035meV$.  
$\chi''= S(1  - e^{-E/k_B T})$ is shown, where the
non-magnetic background, derived from scans with $q$ parallel to the easy
moment direction $c$, has been subtracted. Different colours represent
different temperatures. 
Lines correspond to Eq.(7) by $H=0$ and a temperature independent
$\theta({\bf q})^{\alpha}$ expanded in even powers of {\bf q}.
\label{2}

\item 
{\bf Temperature dependence of the inverse static
susceptibility.} (a) $\chi({\bf q})$ (from direct magnetization
(${\bf q}=0$) and neutron data (${\bf q} \neq 0$)), showing the same anomalous
power-law dependence on temperature, over a wide range of different
momentum values (see inset in fig.2a).  
b) Checking the same anomalous power-law in
temperature of $1/\chi({\bf q}=0)$ for different Au concentrations $x$. The
temperature below which deviations occur increases with distance to
the critical concentration $x=0.1$.
\label{3}

\item  {\bf Energy and field scaling Plots.}(a) $E/T$ scaling plots taken at various critical ${\bf q}$ vectors
(labelled as A and B as in the inset of fig. (2a)), combining previous
triple-axis data with the much more extensive time-of-flight data
taken at IRIS. The line is the product of the Bose-Einstein factor $(1
- e^{-E/k_B T})$ multiplied by the scaling function
$g(y)=C/(1-iy)^{\alpha}$ with $y=E/(aT)$ in eq.(3) with exponent
$\alpha =0.75$. Right inset shows the wide range of $E-T$
covered. Left inset shows the ``scatter'' of the scaling plot as a
function of $\alpha$, being minimal for $\alpha=0.72\pm0.05$.  (b)
H/T-scaling of the local contribution to the uniform magnetization
$M(T,H)$.  $1/\chi_{0}= (dM/dH)^{-1} -(4.1 \mu_B^2/\hbox{mev})^{-1}$.
Solid line corresponds to scaling function $f(h)=(1+h^2)^{-\alpha/2}$
with $h=g\mu_BH/k_BT$ with $\alpha = 0.75$ and an effective moment
$\mu =g\mu_B=1.5 \mu_B$.  Inset shows the $H-T$ range where scaling
applies.
\label{4}

\end{enumerate}
\vfill \eject


\bxwidth=0.5\textwidth
\prk{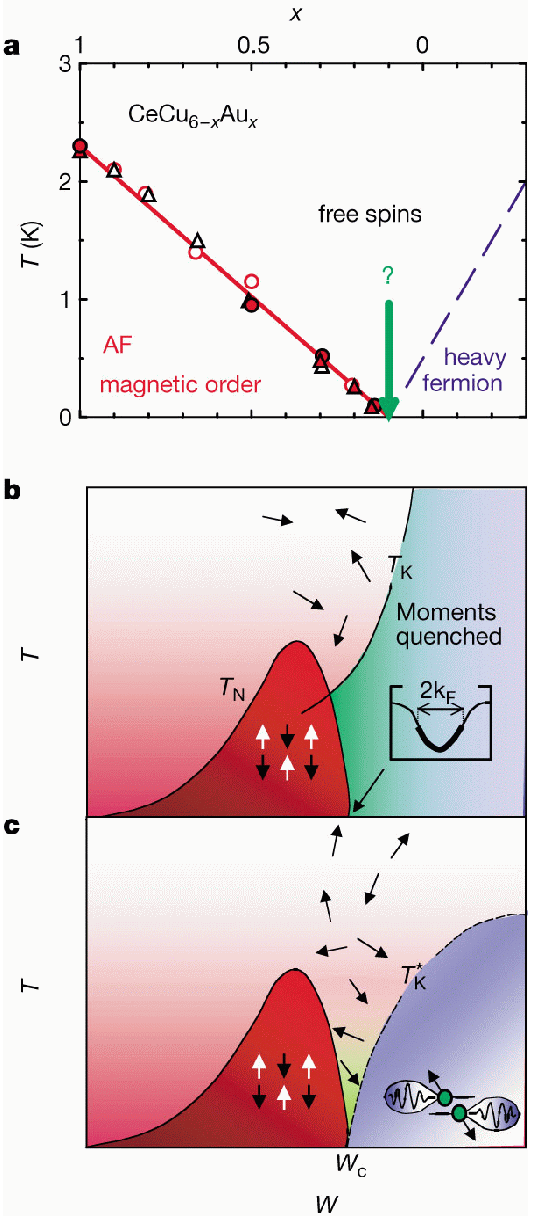}{1}

\newpage
\bxwidth=\textwidth
\prm{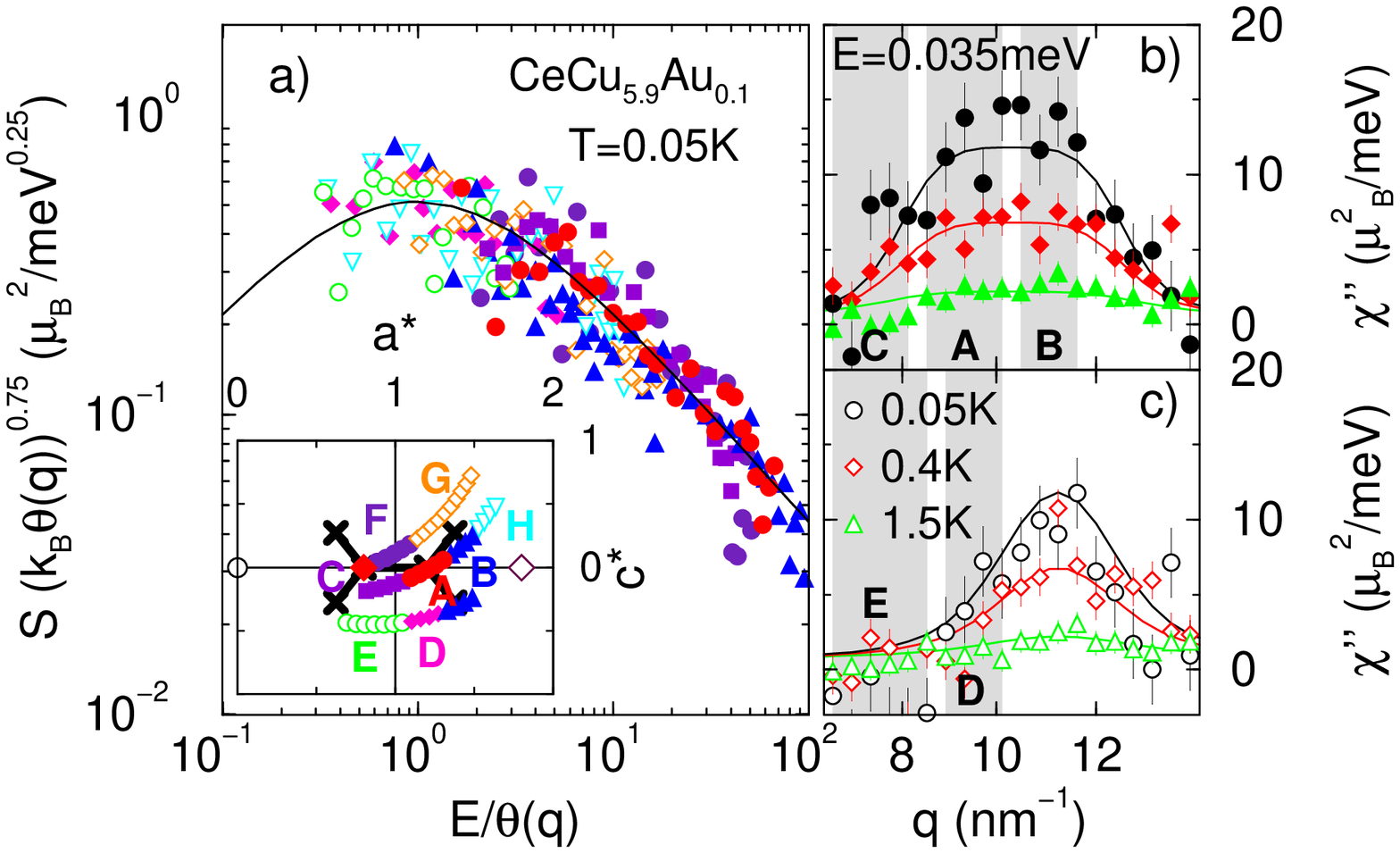}{2}


\newpage
\bxwidth =0.9 \textwidth
\prm{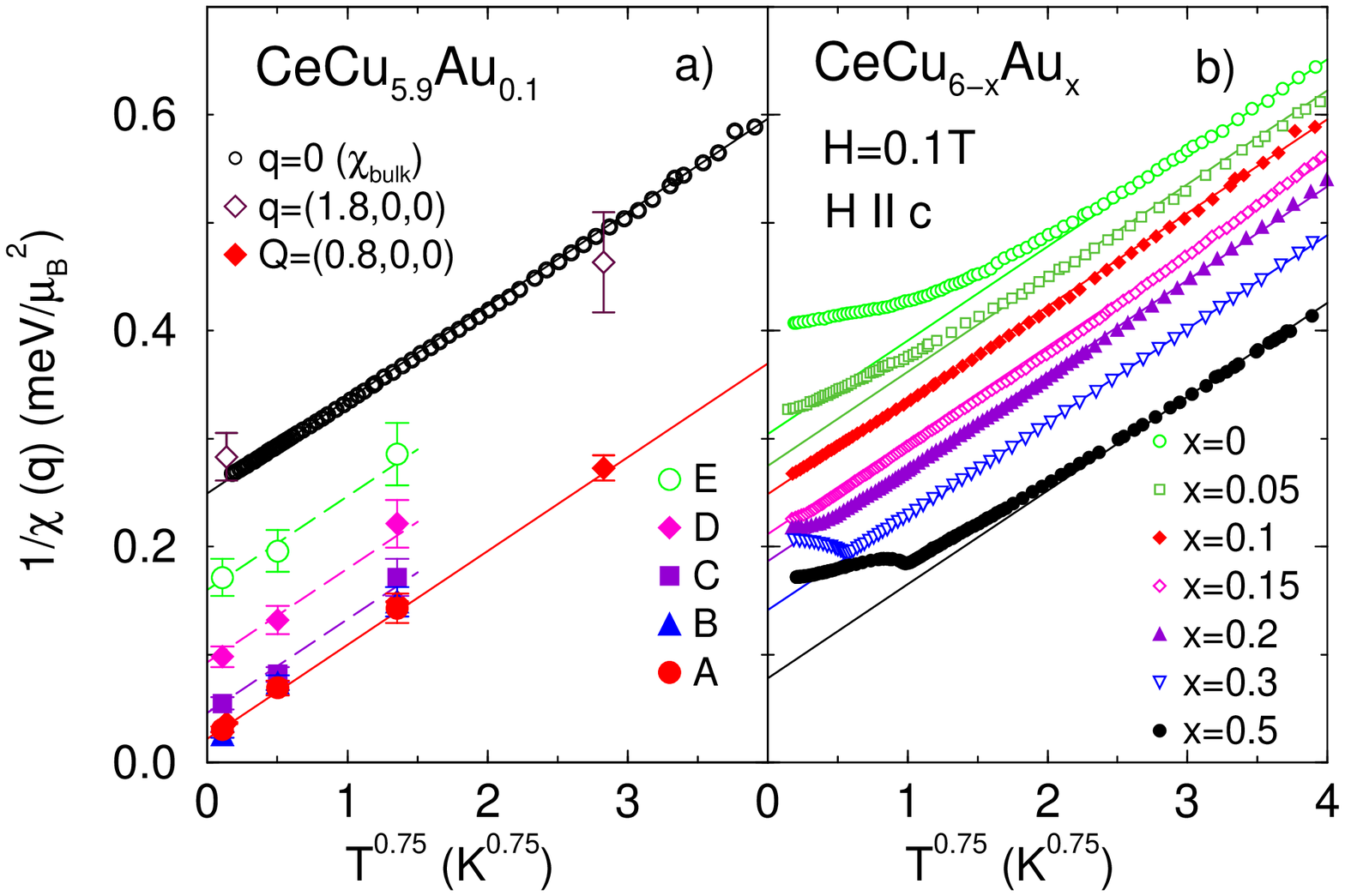}{3}


\prm{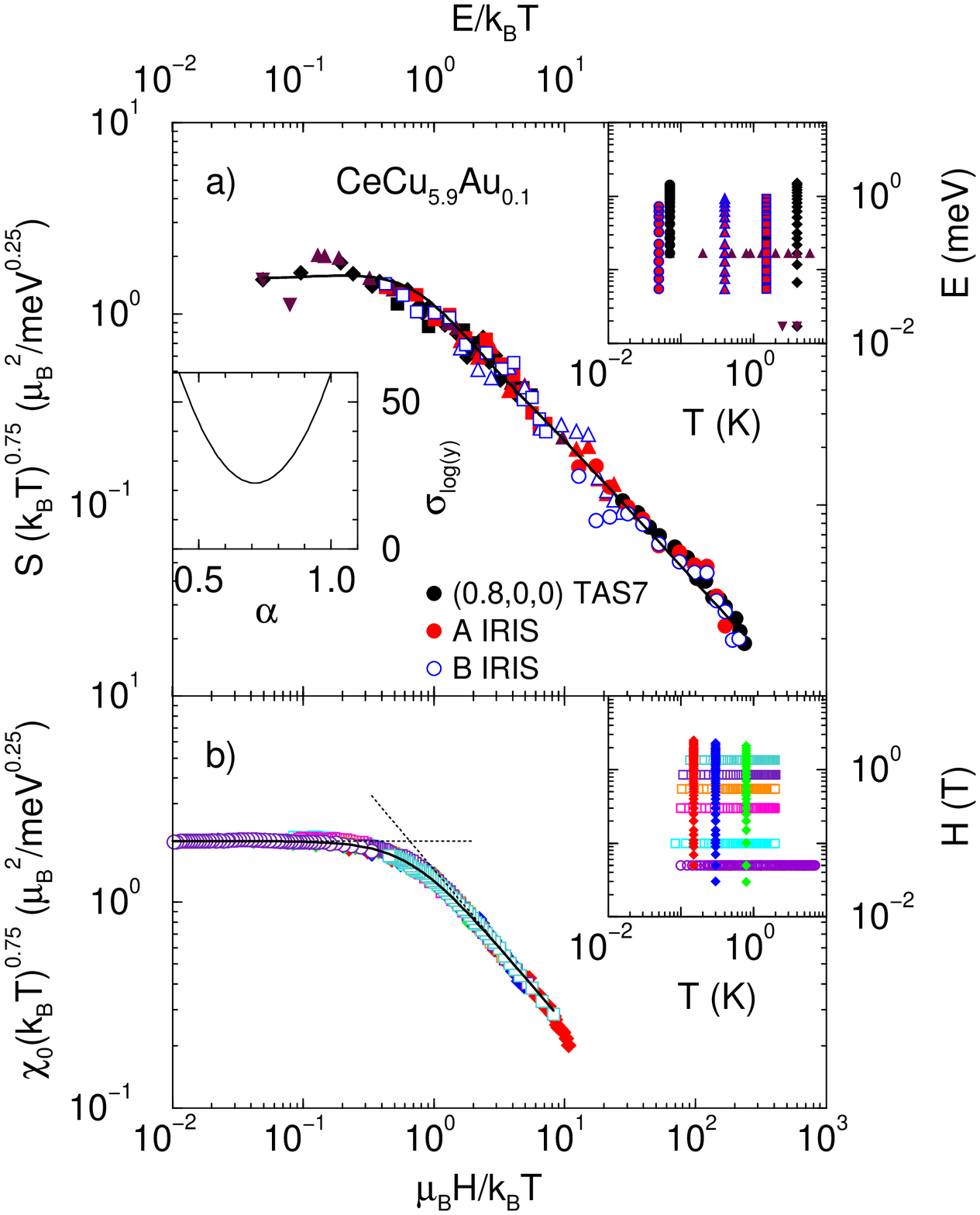}{4}


\end{document}